\documentclass{fic-l}
\usepackage{graphicx}
\usepackage{citesort}
\usepackage[usenames]{color}

\newtheorem{theorem}{Theorem}[section]

\theoremstyle{definition}

\theoremstyle{remark}

\numberwithin{equation}{section}

\newcommand{\be}{\begin{equation}}
\newcommand{\ee}{\end{equation}}
\newcommand{\bea}{\begin{eqnarray}}
\newcommand{\eea}{\end{eqnarray}}

\def\eq#1{(\ref{#1})}

\def\fig#1{fig.~\ref{#1}}
\def\ins#1#2#3{\hskip #1cm \hbox{#3}\hskip #2cm}

\def\e#1{\,{\rm e}^{\displaystyle #1}}
\newcommand{\Z}{ \mathcal{Z}}
\def\phi{ \varphi }
\def\A{\mathcal{A}}
\def\C{\mathcal{ C}}
\def\D{\mathcal{ D}}
\def\F{\mathcal{ F}}
\def\Bb{\bar{B}}
\def\Db{\bar{D}}

\def\rt{{\tilde r}}
\def\rh{{\hat r}}

\newcommand{\inte}{\! \int \!\!}

\newcommand{\ie}{{\it i.e.}\ }
\newcommand{\cf}{{\it cf.}\ }
\newcommand{\eg}{{\it e.g.}\ }

\newcommand{\viz}{{\it viz.}\ }

\newcommand{\fder}[2]{\ensuremath{\frac{\delta #1}{\delta #2}}}
\newcommand{\hf}{\frac{1}{2}}

\def\dd{\dot{\Delta}}
\def\ker#1{\!\cdot\! #1 \!\cdot\!}
\def\hS{\hat{S}}
\def\e#1{\,{\rm e}^{\displaystyle #1}}
\def\one{\hbox{1\kern-.8mm l}}
\def\str{\mathrm{str}}

\newcommand{\tr}{{\mathrm{tr}}}

\newcommand{\tb}[1]{{\tilde\beta}_#1}

\begin{document}

\title{Manifestly Gauge Invariant Exact Renormalization Group
}

\author{Stefano Arnone,}
\address{Dipartimento di Fisica,\\ Universit\`a degli Studi di Roma ``La
Sapienza'', Italy}

\author{Tim R. Morris and Oliver J. Rosten}
\address{School of Physics and Astronomy,\\  University of Southampton, UK}


\dedicatory{Talk presented by TRM at RG2005, Helsinki, Finland,
September 2005 and Renormalization and Universality in
Mathematical Physics Workshop, Fields Institute,  Toronto, Canada,
October 2005, extended to include more details on the strong
renormalized coupling expansion.}

\begin{abstract}
We construct  a manifestly gauge invariant Exact Renormalization
Group for $SU(N)$ Yang-Mills theory, in a form suitable for
calculations without gauge fixing at any order of perturbation
theory. The effective cutoff is incorporated via a manifestly
realised spontaneously broken $SU(N|N)$ gauge invariance.
Diagrammatic methods are developed which allow the calculations to
proceed without specifying the precise form of the cutoff
structure. We confirm consistency by computing for the first time
both the one and two loop beta function coefficients without
fixing the gauge or specifying the details of the cutoff. We
sketch how to incorporate quarks and thus compute in QCD. Finally
we analyse the renormalization group behaviour as the renormalized
coupling becomes large, and show that confinement is a consequence
if and only if the coupling diverges in the limit that all modes
are integrated out. We also investigate an expansion in the
inverse square renormalized coupling, and show that under general
assumptions it yields a new non-perturbative approximation scheme
corresponding to expanding in $1/\Lambda_{QCD}$.
\end{abstract}

\maketitle

\section{Adapting Exact RG to gauge invariant systems}

\subsection{Motivation: obvious advantages}

We want to take the Exact RG (Renormalization Group),\footnote{The
Exact RG was discovered and christened simultaneously by Wegner
and Wilson \cite{Wegner,Wilson}} which encapsulates Wilson and
Fisher's ideas \cite{Wilson,Fisher} about the non-perturbative RG,
adapted to the continuum, and apply these ideas to gauge invariant
systems. We use the continuum version because we are interested in
the quantum field theory description of particle physics. The
reason we want to bring gauge invariance and the Exact RG together
is because each of these ideas are important and powerful in
themselves.

Although the Exact RG was conceived in the 1970s it has only been
understood more recently that it is much more than just a formal
device.  It is a powerful framework for doing computations in
quantum field theory . In brief, the reasons for this are as
follows. (For reviews, see \cite{Fisher,reviews,reviewst,0}. We
will cover some aspects in more detail later.) Firstly, from the
way it is constructed, RG invariance is built in from the
beginning. This means it is particularly easy to define the main
object of interest in quantum field theory, namely the continuum
limit: the Exact RG is an equation that describes how the
Wilsonian effective action changes as the Wilsonian effective
cutoff is changed. To get a continuum limit one simply searches
for a self-similar solution (the ``functional self-similarity'' of
Shirkov \cite{Shirkov}), \ie an effective action whose only
dependence on the cutoff (after writing all quantities as
dimensionless ratios using the cutoff) is through a finite set of
couplings.\footnote{The term ``coupling'' stands for masses also.}
Secondly, all the information one could want can be extracted
directly from the Wilsonian effective action. This is because the
action is effectively the generator of connected Green functions,
which in turn gives the S-matrix elements and thus anything that
can be asked of the quantum field theory. Finally there is a great
deal of freedom in the choice of approximations that can be
applied. Of course the equations can be solved in perturbation
theory, however there is also a limitless variety of
non-perturbative approximations: the exact RG equations can simply
be truncated, or more motivated model approximations can be made
-- adapted to the physics being studied. All these approximations
still preserve the property of RG invariance and existence of
continuum limits. These are the reasons why this framework has
been popular avenue of exploration in the last decade or so
\cite{reviews}.

It would be hard to overemphasise the importance of gauge
invariance to particle physics. It underlies the Standard Model
and attempts to go beyond the Standard Model, and in fact much
more besides than particle physics. Therefore if the ERG framework
is to be useful in particle physics and more generally, we do need
to understand how best to combine it with gauge invariance.

\subsection{The problem}

Given that gauge invariance and the non-perturbative RG are so
important, this problem would have been solved long ago if it were
not for the fact that their combination presents special
difficulties. Indeed at first sight the two concepts are
incompatible. In the Wilsonian RG the first step is a Kadanoff
blocking \cite{Kadanoff} and in the continuum this means
integrating out momentum modes down to some momentum cutoff. But a
momentum cutoff breaks gauge invariance.

There are only two ways to proceed. Either the gauge invariance is
broken by the cutoff or the Exact RG is generalised so as to
incorporate a gauge invariant notion of a cutoff. In the first
alternative it is now well understood how to recover the gauge
invariance at least in principle: the breaking can be quantified
by a set of broken Ward identities, which if satisfied exactly at
some point on the flow\footnote{This is the hard part. It can be
achieved in perturbation theory with some work \cite{Becchi}.}
remain satisfied elsewhere on the flow, and furthermore turn into
the unbroken Ward identities once all momentum modes are
integrated out.

\subsection{Hidden advantages: manifest gauge invariance}

We will explain how to implement the second alternative. In this
case gauge invariance is exactly preserved at all points along the
flow, allowing us to exploit its elegance and power. However, as
we will see, we will need to introduce some extra regularisation
structure.

As well as the obvious benefits of this approach that we have been
describing, it turns out there are a couple of surprise benefits.
Because we have to generalise what we mean by an Exact RG, we are
forced to realise that there are in fact infinitely many different
forms of Exact RG \cite{Jose}. We can use this inherent freedom to
help solve the system, in particular in this case to make the flow
equation itself manifestly gauge invariant. More suprising
however, when we come to solve the flow equation we find that we
do not have to fix the gauge. As a result there are no BRST ghost
fields \cite{BRST}, there are no Gribov problems
\cite{Gribov},\footnote{This is an infamous non-perturbative
problem arising from an incomplete gauge fixing which has as yet
no practical solution.} there is no wavefunction renormalization
for the gauge field, and the expressions are simple and tightly
constrained by the exact preservation of gauge invariance. In
particular, expressions can be built purely from covariant
derivatives. (This should be compared to the much more involved
procedure of using BRST invariance, Lee-Zinn-Justin identities and
so forth \cite{LZJ,book}.)

\subsection{Yang-Mills without gauge fixing}

We will concentrate on $SU(N)$ Yang-Mills. It is defined through
the covariant derivative $D_\mu = \partial_\mu -iA_\mu$, where
$A_\mu\equiv A^a_\mu\tau^a$ is the connection, or gauge field,
contracted into the generators of the $SU(N)$ Lie
algebra.\footnote{For simplicity of exposition, we normalise the
generators to $\tr\,\tau^a\tau^b = \delta^{ab}$, differing from
standard practice and our papers
\cite{0,1,2,sunn,4,aprop,thesis,6,7,twoloops,8,RecursionSoln,ops}.}
The field strength $F_{\mu\nu} = i[D_\mu,D_\nu]$ is just a
commutator of covariant derivatives and the effective action,
being gauge invariant, is just an expansion built out of covariant
derivatives:
\be
\label{g}
S[A](g) = {1\over4g^2}\,\tr\!\int\! F^2_{\mu\nu} + \hbox{higher
dimension operators} + \hbox{vacuum energy}.
\ee
This is an example of a continuum limit solution in this
framework. The $F^2_{\mu\nu}$ term is already dimension four so we
know there is only one coupling $g$; we require a solution $S[A]$
which is a function of the scale only through this coupling. In
the quantum field theory we need to define what we mean by $g$
once we go beyond the classical level. The expansion above acts as
this required renormalization condition: our coupling is defined
to be the coefficient of the $F^2_{\mu\nu}$ term as in \eq{g}.

As we have already emphasised we preserve exactly the local
invariance
\be
\label{gaugetr}
\delta A_\mu = [D_\mu,\omega(x)]
\ee
which implies that there is no wavefunction renormalization, so it
really is the case that only $g$ runs. (The higher dimensional
operators in \eq{g} are all irrelevant and are fixed, computable,
functions of $g$. If wavefunction renormalization were needed we
would have to write $A_\mu = Z^{1/2} A^R_\mu$ where $A^R_\mu$ is
the renormalized field, but in terms of this the invariance
becomes $\delta A^R_\mu = Z^{-1/2}\partial_\mu\omega -
i[A^R_\mu,\omega]$. In other words, it is preserved only if $Z=1$
and $A_\mu=A^R_\mu$. This simple argument fails in the usual
framework only because $\omega$ gets replaced by a ghost field,
leading to a divergent product of quantum fields, which requires
further renormalization.)

\subsection{Hidden advantages: diagrammatic approach} A second
surprise benefit that is basically forced on us is as follows. As
we have already intimated, in order to make everything gauge
invariant, we will have to add quite a bit of extra structure. We
can make a particular choice for this extra regularisation but it
does not seem helpful to do so: when computing Feynman diagrams
this corresponds to a particular choice of vertices and
propagators and there does not appear to be any choice that makes
the integrals easy to do.

It turns out that it is better to specify just a set of schemes
satisfying certain general properties (including of course gauge
invariance). This then leads to vertices and propagators that are
not completely defined but instead satisfy certain properties. The
physics in quantum field theory is encoded in universal quantities
whose values are independent of a particular scheme. Therefore it
ought to be possible to extract these values by manipulating the
diagrams, even though its elements are not completely specified.

It turns out that this is indeed the case. With all these elements
unknown, there is so little freedom in manipulating these
expressions that the procedure is essentially algorithmic. This
computational method also furnishes an automatic check of
universality since the final result is either unique, \ie
independent of the details of the scheme, or it was not universal
after all. Inspired by the fact that this procedure is essentially
algorithmic, Dr. Rosten went on to find directly the solution to
the algorithm \cite{thesis,RecursionSoln}, so although in this
report we will demonstrate how to manipulate these diagrams it is
now understood how to jump directly to the answer.

\section{Generalised Exact RGs}

\subsection{Kadanoff blockings} One way to see that there are infinitely
many different Exact RGs is to start from the fact that there are
infinitely many different Kadanoff blockings. Apart from the fact
that we write this in the continuum the standard definition for a
Kadanoff blocking is
\be
\label{Kad}
\e{-S[\phi]} = \int\D\phi_0\ \delta\Big[ \phi - b[\phi_0]\Big]
      \e{-S_{bare}[\phi_0]}.
\ee
The integration is performed over the bare (or microscopic) field
$\phi_0$ (we take for simplicity a single component real scalar
field) weighted by the Boltzmann factor containing $S_{bare}$, the
microscopic action. The effective field, $\phi$, is related to the
microscopic field through the blocking functional $b_x$. A simple
linear blocking is
\[
b_x[\phi_0] =
       \int_y\! K(x-y)\,\phi_0(y)
\]
for some kernel $K(z)$ which is steeply decaying once
$z\Lambda>1$, $\Lambda$ being the momentum cutoff. It can also be
non-linear, for which there is a huge freedom of choice.

Eqn.~\eq{Kad} is the standard definition because, integrating over
the effective field, we immediately see that the effective
partition function is equal to the microscopic one:
\[
       \Z =  \inte{\D\phi}\e{-S[\phi]} = \int{\D\phi_0}
       \e{-S_{bare}[\phi_0]},
\]
therefore no information has been lost, it has only been recast.
We get an Exact RG by differentiating with respect to $\Lambda$:
\[
           \Lambda{\partial\over\partial\Lambda} \e{-S[\phi]} = - \int_x\frac{\delta}{\delta\phi(x)}
           \int\D\phi_0\
           \delta\Big[ \phi - b[\phi_0]\Big] \Lambda{\partial b_x\over\partial\Lambda}\,
           \e{-S_{bare}[\phi_0]}.
\]
We can interpret the integral as, up to normalization, the rate of
change, $\Psi_x$, of the blocking functional in the measure
provided by \eq{Kad}:
\be
\label{erg}
         \Lambda{\partial\over\partial\Lambda} \e{-S[\phi]} =  \int_x
         {\delta\over\delta\phi(x)}\left(\Psi_x
         \e{-S[\phi]}\right).
\ee
We see that we have a flow equation for each choice of blocking,
or equally we have a flow equation for each choice of $\Psi$.
There are infinitely many such flow equations and by construction
they leave the partition function, $\Z$, invariant. This is
equally clear from \eq{erg}, since $\partial_\Lambda\Z$ is the
integral over $\phi$ of a total derivative with respect to $\phi$,
which vanishes for sensible $S$.

\subsection{Polchinski's exact RG for a single massless scalar field}
Every exact RG can be written in the form \eq{erg}. One of the
most popular is Polchinski's \cite{Joe} for which the rate of
change is
\be
\label{polPsi}
\Psi_x = \hf \int_y \!\dd_{xy}\,
     \fder{\Sigma_1}{\phi(y)}.
\ee
We have introduced what we will call the exact RG kernel,
$\dd\equiv -\Lambda\partial_\Lambda\Delta$. Here and later the
over-dot is the differential with respect to `renormalization
group time'. In this case we are differentiating the `effective
propagator'
\be
\label{prop}
 \Delta = \frac{c(p^2/\Lambda^2)}{p^2}.
\ee
Here at last we make explicit the effective cutoff in the theory:
the cutoff function $c$ is a smooth function such that $c(0)=1$
and $c$ vanishes for large momenta sufficiently rapidly that all
Feynman diagrams are regularised when $\Delta$ is used in place of
$1/p^2$. $\Sigma_1=S - 2\hS$ contains the `seed action' $\hS$,
something that one is free to choose and which for the Polchinski
flow equation just corresponds to what we expect for the
regularised kinetic term $\hS= \hf\partial_\mu\phi\cdot
c^{-1}\cdot\partial_\mu\phi$, yielding the inverse of \eq{prop}.
We will later use this as our starting point for generalisation.
We write, here and later, the momentum kernel contracted into
functions in a number of equivalent ways:
\bea
f \ker{W} g &=& \int_{x,y} f(x)\, W_{xy}\,g(y) = \int_x f(x)
W(-\partial^2/\Lambda^2)\, g(x)\nonumber\\
W_{x y} &=& W(-\partial^2/\Lambda^2)\,\delta(x-y) = \int_p
W(p^2/\Lambda^2) \, {\rm e}^{i p \cdot (x-y)}.\nonumber
\eea

Substituting \eq{polPsi} into the general form \eq{erg}, it easy
to see that we get the following equation for the flow of the
effective action with respect to RG time:
\be
\label{pol}
\dot{S} = {1\over 2} {\delta S\over\delta\varphi} \ker{\dd}
{\delta \Sigma_1 \over\delta\varphi} - {1\over 2}{\delta
\over\delta\varphi} \ker{\dd} {\delta \Sigma_1\over\delta\varphi}.
\ee
\begin{figure}[h!]
\begin{center}
\includegraphics[height=1.7cm]{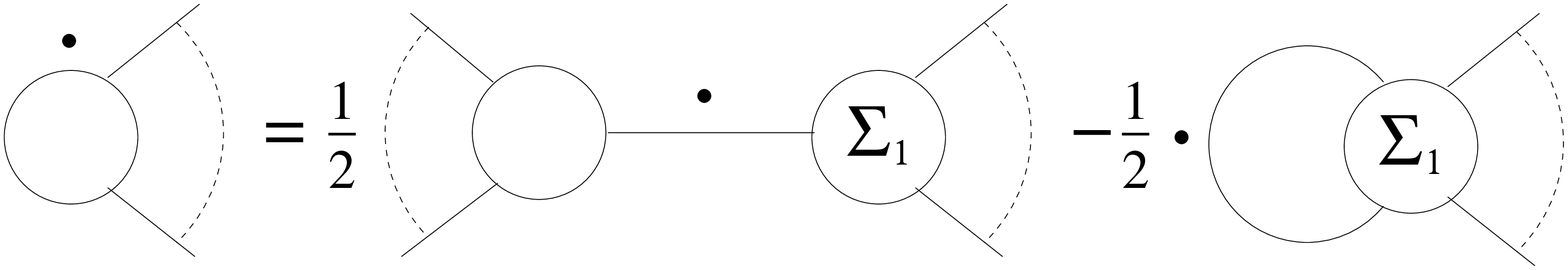}
\end{center}
\caption{Diagrammatic representation of the flow equation for a
single scalar field.}
\label{scaflow}
\end{figure}
The two terms are straightforward to interpret diagrammatically,
as shown in \fig{scaflow}. On the LHS (left hand side) of the
diagram we have a differentiated vertex with some number of legs.
As already intimated, this is constructed from connected
(one-particle-reducible) Feynman diagrams, where instead of the
usual propagator, an effective propagator is used.\footnote{In
fact the effective propagator here is the infrared regulated
propagator $(1-c)/p^2$, with differential $-\dd$, corresponding to
the modes already integrated out \cite{Me1}.} Thus the
differential with respect to RG time either hits a propagator
which divides the diagram in two, giving the first, tree-level
type, diagram on the RHS (Right Hand Side), or it hits a
propagator carrying loop momenta, in which case the result is the
second, one-loop-like, diagram.

Now consider only $S_0$, the classical part of the effective
action. Its flow is governed by just the classical part:
\[
\dot{S_0} = {1\over 2} {\delta S_0\over\delta\varphi} \ker{\dd}
{\delta \Sigma_1 \over\delta\varphi}.
\]
\begin{figure}[h!]
\begin{center}
\includegraphics[height=1.7cm]{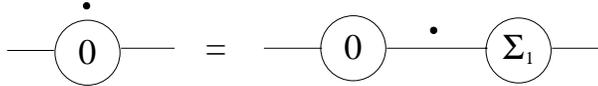}
\end{center}
\caption{The Polchinski flow equation for the classical two-point
vertex}
\label{sca02ptflow}
\end{figure}
If we concentrate just on the two-point vertex, diagrammatically
the flow is of the form of \fig{sca02ptflow}. It turns out that
for the Polchinski equation one can set the classical two-point
vertex equal to the seed action two-point vertex. Then the flow
equation has a straightforward solution:
\be
\label{scaeffprop}
S_0^{\phi\phi}(p)=\hS^{\phi\phi}(p) \implies
\dot{S}^{\phi\phi}_0 = - S_0^{\phi\phi}\dd S_0^{\phi\phi} \implies
\Delta = \left(S_0^{\phi\phi}\right)^{-1},
\ee
so $\Delta$ is just the inverse of the classical effective action
two-point vertex, justifying our naming it an effective
propagator.

Later we will turn this analysis around, and insist that the
classical and seed action two-point vertices can be identified and
then the equation above determines what we use for the exact RG
kernel, or equivalently the effective propagator. This step is not
really necessary but it does lead to technical simplifications.

\subsection{Generalised exact RG for a single scalar field} It is
now straightforward to generalise this. For example we can replace
the seed action by $\hS= \hf\partial_\mu\phi\cdot
c^{-1}\cdot\partial_\mu\phi+\cdots$, where the ellipses stands for
some higher point vertices, or even an infinite number of higher
point vertices. The only constraint, apart from the usual generic
requirements, is that these vertices' ultraviolet behaviour should
not be so violent as to destroy the regularisation properties of
$c$. In fact since the two-point vertex is the most general
two-point vertex consistent with there being only one scale,
$\Lambda$, and Poincar\'e invariance, in effect we replace the
seed action by {\it any effective action of our choice}, as long
as it corresponds to a regularised massless scalar field.

What changes? Hardly anything! The equation \eq{pol} and its
diagrammatic interpretation, \fig{scaflow}, are still the same.
The only difference is that one must remember that $\hS$ vertices
can now have more than two legs. It might seem surprising but the
universal answers extracted from this equation are exactly the
same as in standard quantum field theory. We have checked this
explicitly. For example, for a single scalar field in four
dimensions the first two beta function coefficients come out
exactly the same, independent of the details of $c$ and the higher
point vertices \cite{sca12}. The underlying reason for this is
that these extra terms amount to some reparametrisation of the
field as the cutoff is lowered \cite{Jose}, and of course such a
change of field variables does not change the physics.

\section{Manifestly gauge invariant Exact RG for Yang-Mills}

\begin{figure}[h!]
\begin{center}
\includegraphics[height=1.7cm]{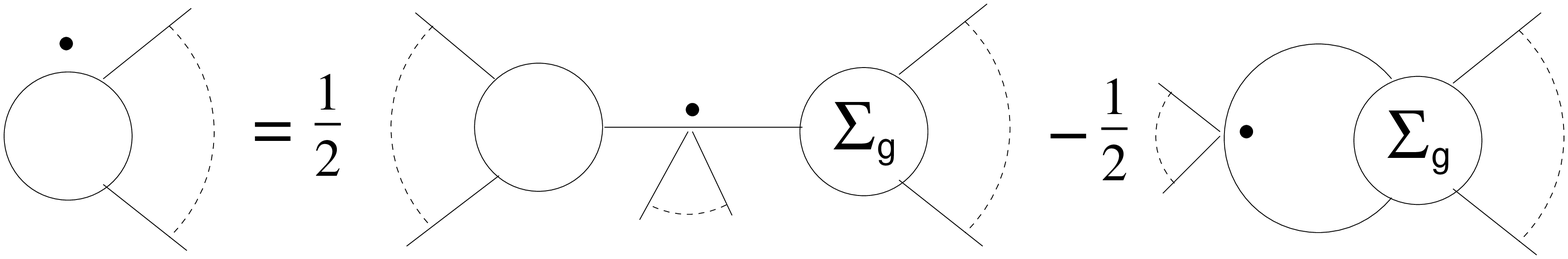}
\end{center}
\caption{The flow equation in gauge theory.}
\label{fAflow}
\end{figure}

Replacing the scalar field $\phi$ by a gauge field $A_\mu$, we can
just require that the seed action is any choice of effective
action: we do not even have to specify that the field is massless,
since gauge invariance will ensure that. However making these
simple replacements clearly does not turn \eq{pol} into a gauge
invariant equation, because the gauge invariance is broken by the
kernel. We can solve this problem by replacing
$\dd(-\partial^2/\Lambda^2)$ by $\dd(-D^2/\Lambda^2)$. This is
only one of infinitely many ways to covariantize a kernel. We will
let braces stand for some fixed choice, which we will never have
to specify precisely. Now the flow equation is gauge invariant:
\be
\label{Aflow}
\dot{S} = {1\over 2} {\delta S\over\delta A_\mu} \{\dd\}
{\delta \Sigma_{g} \over\delta A_\mu} - {1\over 2}{\delta
\over\delta A_\mu} \{\dd\} {\delta \Sigma_g \over\delta A_\mu}.
\ee
It corresponds to the gauge covariant rate of change:
\[ \Psi = \hf\, \{\dd\}\fder{\Sigma_g}{A_\mu}. \]
At the diagrammatic level, the only difference is that we have to
remember that there are now fields hidden in the kernels, so the
kernels themselves have vertices, as illustrated in
\fig{fAflow}.\footnote{Generically there are also diagrams where
$\delta/\delta\A_\mu$ acts on $\{\dd\}$, which can be excluded by
a careful choice of covariantization \cite{0,aprop}.} One last
detail: we scale the coupling constant out of the action, as in
\eq{g}, and some thought shows that it ends up in front of the
action so that in place of $\Sigma_1$ we have $\Sigma_g =
g^2S-2\hS$.

Having scaled $g$ outside the action, we expect it to count powers
of $\hbar$, and it is easy to check that \eq{Aflow} has this form
for its weak coupling expansion:\footnote{In practice it is
convenient to let $\hS$ also have an $\hbar$ expansion
\cite{thesis,twoloops}.}
\begin{eqnarray*}
  S &=& {1\over g^2} S_0 + S_1 + g^2 S_2 + \cdots \\
  \beta &:=& \Lambda\partial_\Lambda g = \beta_1 g^3 + \beta_2 g^5 +
  \cdots,
\end{eqnarray*}
where $S_n$ and $\beta_n$ are the $n$-loop pieces of the effective
action and $\beta$ function respectively. Of course these
components have yet to be computed.

\subsection{Consequences at the classical level}
\begin{figure}[h!]
\begin{center}
\includegraphics[height=1.7cm]{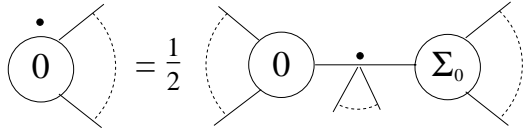}
\end{center}
\caption{The classical flow equation for the classical part of the
effective action.}
\label{fA0flow}
\end{figure}

Classical effective action vertices follow from the $g\to0$ limit
of \eq{Aflow}, which is easily seen to be
\[
\dot{S}_0 = {1\over 2} {\delta S_0\over\delta A_\mu} \{\dd\}
{\delta \Sigma_0 \over\delta A_\mu},\ins11{where} \Sigma_0 =
S_0-2\hS.
\]
Diagrammatically, this takes the form in \fig{fA0flow}.

Recall that we determine the kernel by insisting on the equality
of the classical and seed action two-point vertices. The classical
vertex has the following general form
\[
S_{0\,\mu\nu}^{\phantom{0\,}AA} =
       \Box_{\mu\nu}(p)/c(p^2/\Lambda^2),
       \ins11{where} \Box_{\mu\nu}(p) = \delta_{\mu\nu} p^2 - p_\mu
       p_\nu.
\]
This form is fixed uniquely by masslessness, gauge invariance and
Poincar\'e invariance. The arbitrary function $c$ will again play
the r\^ole of a cutoff function. $\Box_{\mu\nu}(p)$ is the
standard transverse kernel in the kinetic term of a gauge field.
(From here on we suppress its dependence on $p$.) Since it has a
longitudinal zero-mode: $\Box_{\mu\nu}p_\nu=0$, it is not
invertible and this is the standard obstruction in gauge theory
that motivates gauge fixing.

Following our procedure for determining the Exact RG kernel we
see, using $\Box_{\mu\nu}\Box_{\nu\lambda} = p^2
\Box_{\mu\lambda}$, that we get precisely the same solution for
$\dd$ and thus also $\Delta$, as before:\footnote{We do not yet
see the vertices from covariantizing the kernel because that would
require one-point vertices for the gauge field which are not
allowed, by Lorentz invariance for example.}
\[
S_{0\,\mu\nu}^{\phantom{0\,}AA} = \hS^{AA}_{\mu\nu}\ \implies\
\dot{S}_{0\,\mu\nu}^{\phantom{0\,}AA} = -
S_{0\,\mu\alpha}^{\phantom{0\,}AA}\dd
S_{0\,\alpha\nu}^{\phantom{0\,}AA}\ \implies\ \Delta = c/p^2.
\]
This satisfies
\be
\label{Aeffprop}
\Delta S_{0\,\mu\nu}^{\phantom{0\,}AA}
       = \delta_{\mu\nu} - p_\mu p_\nu/p^2.
\ee
Although we will keep calling $\Delta$ an effective propagator, it
is not really an effective propagator any more because the
two-point vertex has no inverse. Instead of unity on the RHS
above, we get the transverse projector, or equivalently
$\delta_{\mu\nu}$ minus the purely longitudinal term $p_\mu
p_\nu/p^2$. This latter term when multiplying other elements in a
diagram can be simplified by using the Ward identities that follow
from invariance under small gauge transformations.

\begin{figure}[h!]
\begin{center}
\includegraphics[height=3cm]{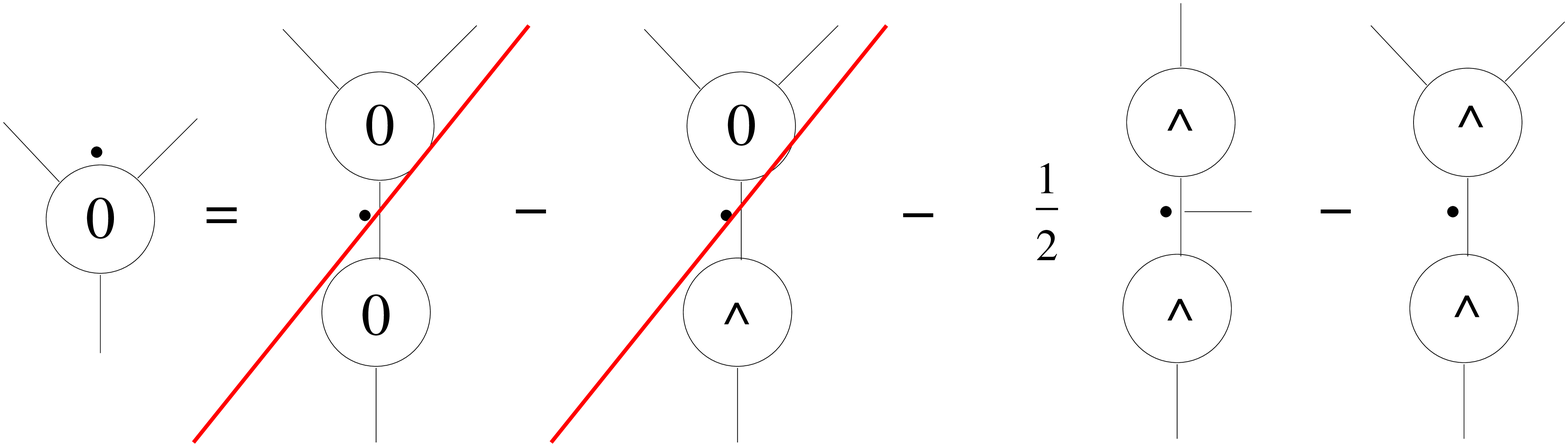}
\end{center}
\caption{The flow equation for the classical three-point vertex.
The seed-action vertices have a caret in the centre.}
\label{fA03flowX}
\end{figure}

Specialising \fig{fA0flow} to the three-point vertex we get the
diagrams in \fig{fA03flowX}.
We now see the effect of covariantizing the kernel, in the third
term on the RHS. The first two terms have the classical
three-point vertex on the top, but they cancel because of the
equality of classical and seed action two-point vertices. The
remaining terms depend only on the covariantized kernel and seed
action vertices, \ie on terms that we put in by hand. Thus
everything is known on the RHS and it easy to integrate this with
respect to RG time to get the full classical three-point vertex.
Notice that we have computed this without having to try to invert
the (non-invertible) kinetic term.

\begin{figure}[h!]
\begin{center}
\includegraphics[height=3cm]{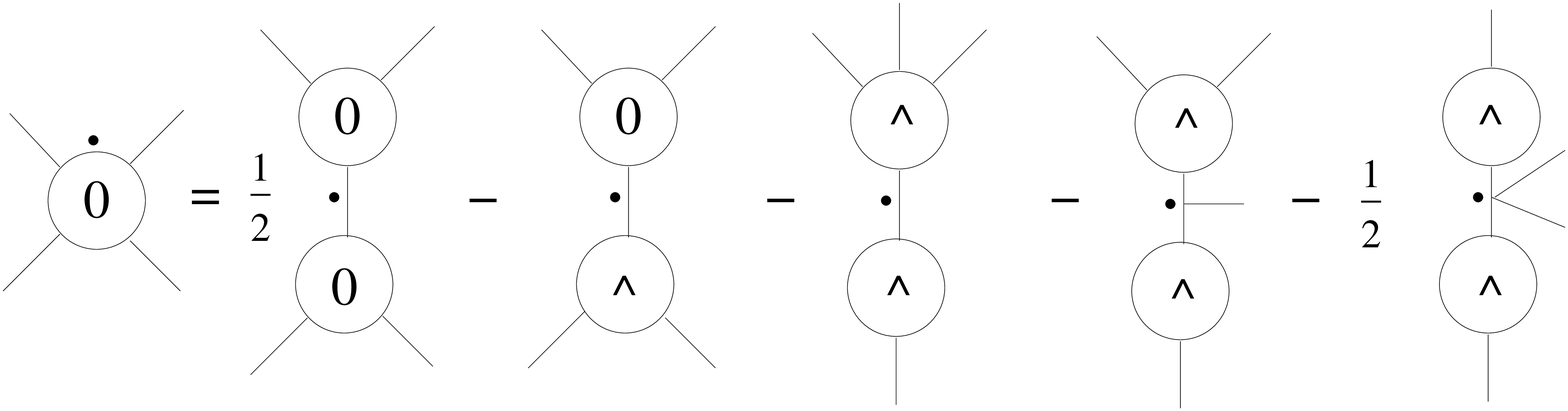}
\end{center}
\caption{The flow equation for the classical four-point vertex.}
\label{fA04flow}
\end{figure}

This cancellation works for all the other vertices at the
classical level: the flow is only into terms that have already
been determined. For example, the flow of the four-point vertex is
shown in \fig{fA04flow}.
On the RHS there are non-vanishing terms with effective action
vertices this time, but these are the three-point vertices that we
have just determined. Therefore this equation is straightforward
to integrate, giving the four-point vertex. In the same way we can
determine all the vertices at the classical level.

\subsection{One loop} 
\begin{figure}[h!]
\begin{center}
\includegraphics[height=2.5cm]{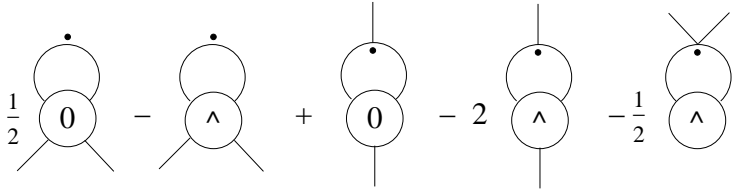}
\end{center}
\caption{Two-point vertex at one loop and $O(p^2)$.}
\label{oneLoop}
\end{figure}

Let us illustrate what happens at one loop
by considering just one particular piece -- the part that yields
the famous asymptotically free $\beta$ function for the coupling
constant $g$. Recall that it is defined through the
renormalization condition \eq{g}. This tells us that the two-point
vertex $S_{\mu\nu}(p)=\Box_{\mu\nu}/g^2+O(p^4)$, where the higher
powers of momentum come exclusively from the higher dimension
operators. Therefore when we compute the flow of $S_{\mu\nu}(p)$
at $O(p^2)$, all we pick up is the flow of $g$, \ie the $\beta$
function. At one loop, from \eq{Aflow}, we find
\be
\label{beta1}
-2\beta_1\Box_{\mu\nu}(p)+O(p^4) = \left.\hf\frac{\delta}{\delta
  A_\alpha}\{\dd\}\frac{\delta\Sigma_0}{\delta
  A_\alpha}\right|^{AA}_{\mu\nu},
\ee
which, when expanded, gives the diagrams of
\fig{oneLoop}.
However, when we compute these diagrams we find that the integrals
do not converge. The ultraviolet divergences are not regulated.
This is not a surprise. Covariantizing the cutoff function in the
way that we have done is equivalent to implementing a covariant
higher derivatives regularisation and this is known to fail,
precisely at one loop \cite{jean}.

We have to replace this effective covariant higher derivatives
with some other gauge invariant regularisation which can naturally
be incorporated in this Exact RG framework, such that it cures
this problem. One way to proceed is to add covariant Pauli-Villars
fields \cite{slavnov} in such a way to as to cancel the leading
divergences. This leads to further divergences which one cancels
by introducing new regulator fields and interactions. Continuing
in this way one finds eventually that the whole framework closes;
the interactions are essentially unique and once in place, all
diagrams, including the new ones, are regularised \cite{0}.

After staring at the result for a while, one realises that there
is a much more elegant way to arrive at the same answer
\cite{2,sunn,4,aprop}.

\section{$SU(N|N)$ regularisation}
\subsection{$SU(N|N)$ Yang-Mills}

Take the $SU(N)$ Yang-Mills and embed $SU(N)$ in the supergroup
$SU(N|N)$. Its Lie algebra is realised by supermatrices, which we
employ in the defining representation. The gauge field becomes a
supergauge field,
\[
\A_\mu =  \left( \!\! \begin{array}{cc}
                   A^{1}_{\mu} & B_{\mu} \\
                   \bar{B}_{\mu} & A^{2}_{\mu}
                   \end{array} \!\!
            \right) + {\A}^{0}_{\mu} \one,
\]
where $A^1$ and $A^2$ are traceless Hermitian matrices and the
off-diagonal terms $B$ and $\Bb$, are Grassmann valued, and thus
anticommuting (fermionic). It turns out we also have to add the
$\A^0$ term, proportional to the $2N\times2N$ unit matrix.

$A^1$ gauges the original $SU_1(N)$ group. We add an index to the
group to distinguish it from an extra copy, $SU_2(N)$, gauged by
$A^2$. The bosonic subgroup is therefore $SU_1(N)\otimes
SU_2(N)\otimes U(1)$, where the $U(1)$ factor is carried by
$\A^0$. If we had been dealing with a compact Lie algebra, as is
normally used in gauge theory, we would be able to discard this
$U(1)$ factor, but we cannot here because the $SU(N|N)$
superalgebra is an example of an indecomposable superalgebra;
$\A^0$ gauges a central term produced by some of the fermionic
generators, for example:\footnote{In a Lie superalgebra, fermionic
generators obey anticommutation laws amongst themselves.}
\[
\left\{\left(\begin{array}{cc} 0 & \one \\ \one & 0 \\
         \end{array}\right), \left(\begin{array}{cc} 0 & \one \\
         \one & 0 \\ \end{array}\right)\right\} = 2\one.
\]
This leads to some technicalities which are nevertheless key, so
we briefly describe them. To form invariants we must use the
supertrace
\be
\label{str}
\str\left(\begin{array}{cc} X^{11} & X^{12} \\
           X^{21} & X^{22} \\ \end{array}\right) = \tr_1 X^{11} -
           \tr_2 X^{22},
\ee
because it is this that is cyclically invariant for supermatrices
and thus this which leads to the construction of $SU(N|N)$
invariants. The action will thus be built from a supercovariant
derivative $\nabla_\mu = \partial_\mu - i \A_\mu$ as
\be
\label{SS}
S\sim {1\over4g^2}\ \str\!\int\!\F^2_{\mu\nu}.
\ee
This has a neat effect on the central term because the kinetic
term for $\A^0$ will just be a $U(1)$ field strength squared times
$\str\one$. But from \eq{str}, $\str\one=N-N=0$. Therefore $\A^0$
has no kinetic term. Furthermore, its interactions are formed from
commutators, but $\one$ commutes with everything and thus $\A^0$
has no interactions. It therefore does not appear at all in the
action displayed above.

We now raise this to a principle and require that the theory be
invariant under the local shift symmetry
\be
\label{noA0}
\delta\A^0_\mu = \lambda_\mu(x).
\ee
This ``no-$\A^0$ symmetry'' simply ensures that nothing in fact
depends on $\A^0$. It means that in practice we gauge the coset
space $SU(N|N)/U(1)$. 

The $SU(N|N)/U(1)$ theory has very nice properties at high
momenta. In particular one can show that the quadratic Casimir
vanishes, which it turns out is sufficient to ensure that the
effective action is finite at one loop. At higher loops, wherever
the $SU(N)$ theory yielded $\tr\one=N$, one now obtains
$\str\one=0$, thus in the 't Hooft large $N$ limit this theory has
no quantum corrections at all.

Note that there is no sense in which this can be regarded as a
physical theory, in particular the $B_\mu$ are fermionic integer
spin particles which thus violate Pauli's spin statistics theorem
\cite{Pauli}. The kinetic term for $A^2$ has the wrong sign, as
follows from \eq{SS} and \eq{str}. This leads to negative norm
states in the Fock space \cite{sunn}. However we will be able to
ensure that these sicknesses are felt by the physical $SU(N)$
Yang-Mills sector only at the cutoff scale.

\subsection{$SU(N|N)$ $\rightarrow$ $SU_1(N)\otimes SU_2(N)$
spontaneous breaking} 

At high energies this more symmetric theory has some very useful
ultraviolet properties. At low energies we want to recover just
the $SU(N)$ Yang-Mills. In particle physics we know how to go from
one regime to the other: we use the Higgs mechanism. We therefore
introduce a super-Higgs field, the superscalar
\[
\C = \left( \!\! \begin{array}{cc}
                   C^1 & D \\
                   \bar{D} & C^2
                   \end{array} \!\!
          \right),
\]
the $C^i$ being bosonic and the off-diagonal $D$ being fermionic.
We can arrange a potential so that $\C$ picks up the expectation
value\footnote{Note that $\C$ is actually a $U(N|N)$ adjoint
field, the supertraceful generator being $\sigma_3$. This
nevertheless transforms under $SU(N|N)$, another peculiarity
compared to compact bosonic groups \cite{sunn}.}
\be
\label{vev}
\langle\C\rangle = \Lambda\left(\begin{array}{cc} \one & 0 \\ 0 & -\one \\
         \end{array}\right).
\ee
It is easy to show that this Pauli $\sigma_3$-matrix form breaks
spontaneously all the fermionic directions, and only the fermionic
directions. Thus, via the Higgs mechanism, the $B$s eat the $D$s
and they become massive with masses of order the cutoff, where we
will ultimately be able to ignore them. The `physical' Higgs
fields $C^i$ can also be given masses at the cutoff scale, so all
that remains at low energies are the two massless gauge fields
$A^1$ and $A^2$. However in this gauge invariant approach, because
the gauge fields are charged under different groups, the lowest
dimension interaction between them is
\[
{1\over\Lambda^4}\ \tr_1\left(F^1_{\mu\nu}\right)^2
    \tr_2\left(F^2_{\mu\nu}\right)^2.
\]
This is already dimension 8, so by the
Symanzik-Appelquist-Carazzone theorem \cite{apple} the two sectors
decouple into a direct product in the low energy limit.

At finite $N$, and with this extra $\C$ sector, we need to supply
some covariantized cutoff functions in order to make the theory
ultraviolet finite. It is then possible to show this to all orders
in perturbation theory. For completeness we give the precise
statement and sketch the steps necessary to prove it \cite{sunn}.

\begin{theorem}
Let
\[
S =
  \str\int{1\over4g^2}\F_{\mu\nu}c^{-1}\left(-{\nabla^2\over\Lambda^2}\right)\F_{\mu\nu}
  +\hf\nabla_\mu\C\,{\tilde
  c}^{-1}\left(-{\nabla^2\over\Lambda^2}\right)\nabla_\mu\C
  +{\lambda\over4}\left(\C^2-\Lambda^2\right)^2,
\]
where $c^{-1}$ is a polynomial of rank $r$ and ${\tilde c}^{-1}$
  is a polynomial of rank ${\tilde r}$. Shifting
  $\C\mapsto\C+\Lambda\left(\begin{array}{cc} \one & 0 \\ 0 & -\one \\
  \end{array}\right)$, all amplitudes are finite to all
  orders in perturbation theory providing $r>{\tilde r}+1$ \&
  ${\tilde r}>1$.
\end{theorem}

We use standard techniques of perturbative field theory for the
proof, to avoid confusing the issue of regularisation with those
of manifest gauge invariance and the Exact RG description. The
first step therefore is to gauge fix to 't Hooft gauge. We
introduce a further cutoff function ${\hat c}^{-1}$ (rank $\rh$)
for the gauge fixing term and thus also the BRST ghosts. Providing
that $\rh\ge r>\rt-1$ and $\rt>-1$ the leading contributions at
high momentum have the symmetries of the symmetric regime. By
power counting, one can show that if $\rh\ge r$, $r-\rt>1$ and
$\rt>1$, all diagrams are superficially finite except certain
``remainder contributions": symmetric phase parts of one-loop
graphs with no external $\C$ or ghost lines and up to four
external $\A$s. Of these, two-point and three-point diagrams
vanish by properties of the superalgebra since one can show that
they involve only the following forms: $\str\A\,\str\A$,
$\str\A\A\,\str\one$, $\str\A\A\,\str\A$ and $\str\A^3\,\str\one$.
These all vanish because $\str\one=0$ and $\str\A=0$. This leaves
only the symmetric phase four-point diagram which can be proved to
be finite by Lee-Zinn-Justin identities \cite{LZJ}. The result is
only conditionally convergent. This can be made well defined by an
appropriate limiting procedure. Defining the theory in
dimension $D\to4^-$ is sufficient. 

\section{Manifestly $SU(N|N)$ gauge invariant Exact RG} We now
combine the ideas in the last two sections. Given the field
content, this amounts to finding an appropriate choice for $\Psi$
in \eq{erg}.\footnote{For technical reasons it is helpful to make
$\C$ dimensionless \cite{1,2,aprop}. For simplicity of exposition
we do not do this in what follows.}
\begin{eqnarray}
  \dot{S} &=& {1\over 2} {\delta S\over\delta \A_\mu} \{\dd^\A\}
{\delta \Sigma_g \over\delta \A_\mu}
    + {1\over 2} {\delta S\over\delta \C} \{\dd^\C\} {\delta \Sigma_g \over\delta
    \C} \label{gierg}\\
  &+& {1\over 2} [\C,{\delta S\over\delta \A_\mu}] \{\dd^\A_m\}
    [\C,{\delta \Sigma_g \over\delta \A_\mu}]
    + {1\over 2} [\C,{\delta S\over\delta \C}] \{\dd^\C_m\}
    [\C,{\delta \Sigma_g \over\delta \C}]\nonumber \\
  &+& {1\over 2} {\delta S\over\delta \A_\mu} \{\dd^\A_\sigma\}
    \left(\{\C,{\delta \Sigma_g \over\delta \A_\mu}\}\,\str\,\C
    - 2\C\,\str\left\{\C{\delta \Sigma_g \over\delta \A_\mu}\right\}\right)\
    + \left(S\leftrightarrow\Sigma_g\right)\nonumber \\
    &+&\  \hbox{quantum terms}.\nonumber
\end{eqnarray}
On the first line we have the term expected already which gives
the flow equation for the gauge (now supergauge) field, together
with its kernel which we have yet to determine. We also have a
piece for the super-Higgs and its corresponding kernel. The whole
flow equation must be invariant under
\be
\label{gauge}
\delta\C = -i[\C,\Omega],\quad \delta\A_\mu = [\nabla_\mu,\Omega]
+ \lambda_\mu\one,\quad
\delta\!\left({\delta\over\delta\A_\mu}\right) =
    -i\left[{\delta\over\delta\A_\mu},\Omega\right]+{\tilde\lambda}_\mu\one,
\ee
\ie the combination of $SU(N|N)$ supergauge transformations, and
the no-$\A^0$ symmetry that ensures we divide out by the central
term.\footnote{The natural definition of $\delta/\delta\A$ is
traceless. This determines ${\tilde\lambda}$ \cite{aprop}.} The
top line is already invariant under these but it is not enough
because we want to keep the technical simplifications that follow
from setting the classical two-point vertex equal to the seed
action two-point vertex. This means that in the spontaneously
broken phase we have to allow the kernels for the fermionic fields
to grow a mass term, because the corresponding effective
propagators must be inverses of their now-massive two-point
vertices. The simplest such terms are displayed on the second line
of \eq{gierg}: using \eq{vev}, the commutators isolate (at the
two-point level) just the fermionic directions as required. We are
not quite done yet. We have to remember that really we now have
two coupling constants, one for our original gauge field $A^1$ and
one for the unphysical copy $A^2$. At the quantum level we can no
longer identify the two couplings because the wrong sign action
for $A^2$ results in its coupling running with the wrong sign. (It
is asymptotically trivial rather than asymptotically free.) In the
way we are doing things, this changes the normalisation of the
$A^2$'s kinetic term. Thus, if we are to keep the equality of
$\hS$ and $S_0$ two-point vertices we have to allow for different
effective propagators for the two gauge fields. The simplest term
that allows for this in the flow equation, whilst still obeying
the symmetries \eq{gauge}, is the multi-supertrace term on the
third line of \eq{gierg}. Finally the quantum terms are just
analogous to the last term in \eq{Aflow}, as determined by the
general form \eq{erg}.

(One can further understand the appearance of these
multisupertrace terms as follows. Classically the effective action
contains the terms:
\[
S_0 = {1\over4g^2}\ \str\int
  \left(\begin{array}{cc} \left(F^1_{\mu\nu}\right)^2 & 0 \\ 0 &
  \left(F^2_{\mu\nu}\right)^2 \\
  \end{array}\right) + \cdots.
\]
Since the one-loop correction has opposite sign for $A^2$, it
takes the form
\[
-{\beta_1\over g^2}\ \ln{\Lambda\over\mu}\ \str\left(\begin{array}{cc} \one & 0 \\ 0 & -\one \\
         \end{array}\right)\left(\begin{array}{cc}\left(F^1_{\mu\nu}\right)^2 & 0 \\ 0 &
  \left(F^2_{\mu\nu}\right)^2 \\
  \end{array}\right).
\]
At first sight such a correction is a puzzle since na\"\i ve
guesses for the full contribution as a functional of $\A$, do not
obey the no-$\A^0$ symmetry. In fact, as we confirmed by direct
calculation, this term arises from a contribution of the form
\[
\str\,\C\, \str\,\C\F_{\mu\nu}^2 -
  \left(\str\,\C\F_{\mu\nu}\right)^2|_{\C\to\C+\Lambda\sigma_3}
  +\cdots,
\]
where the ellipses stands for higher dimension invariant terms.
This does obey no-$\A^0$ symmetry as can be seen by using
$\delta\F_{\mu\nu} =
  (\partial_\mu\lambda_\nu-\partial_\nu\lambda_\mu)\one.$
The terms we added in the flow equation have a similar structure
and indeed allow contributions such as above to be incorporated at
tree-level, as now required.)

\subsection{The kernels} The kernels in \eq{gierg} are determined in the same
way as before. We first write down the classical two-point
vertices in a form as general as required and such that they are
consistent with broken supergauge invariance and the
renormalization conditions (and Poincar\'e invariance {\it etc.}).
%
We then insist that the $\hS$ two-point vertices are equal to
these and this determines the kernels. For example, the two-point
classical flow equations for the gauge fields take the form
\[
\dot{S}_{0\,\mu\nu}^{\phantom{0\,}A_1A_1} = -
       S_{0\,\mu\alpha}^{\phantom{0\,}A_1A_1}\dd^1
       S_{0\,\alpha\nu}^{\phantom{0\,}A_1A_1}\ins11{and}
       \dot{S}_{0\,\mu\nu}^{\phantom{0\,}A_2A_2} = -
       S_{0\,\mu\alpha}^{\phantom{0\,}A_2A_2}\dd^2
       S_{0\,\alpha\nu}^{\phantom{0\,}A_2A_2},
\]
determining $\Delta^1$ and $\Delta^2$, which are the linear
combinations $\Delta^1 = \Delta^\A + 8N\Delta^\A_\sigma$ and
$\Delta^2 = \Delta^\A - 8N\Delta^\A_\sigma$. In the same way, the
other two-point classical flow equations determine linear
combinations of the original kernels, and inverting we finally
find expressions for the original kernels. The superscalar field
classical two-point vertex flows in the same way as the scalar
field considered earlier:
\[
\dot{S}_0^{CC} = - S_0^{CC}\dd^C S_0^{CC}.
\]
However, the $B$ and $D$ fields mix which other. This is just the
Higgs effect in this manifestly gauge invariant description. We
therefore use $2\times2$ matrices to describe the flow:
\[
\left(\!\!
\begin{array}{cc}
  \dot{S}_{0\,\mu\nu}^{\phantom{0\,}\Bb B} & \!\!\!\dot{S}_{0\,\mu}^{\phantom{0\,}\Bb D} \\
  -\dot{S}_{0\,\nu}^{\phantom{0\,}\Bb D} & \!\!\!-\dot{S}^{\Db D}_0 \\
\end{array}\!\!
\right) = - \left(\!\!\begin{array}{cc}
  S_{0\,\mu\alpha}^{\phantom{0\,}\Bb B} & \!\!\!S_{0\,\mu}^{\phantom{0\,}\Bb D} \\
  -S_{0\,\alpha}^{\phantom{0\,}\Bb D} & \!\!\!-S^{\Db D}_0 \\
\end{array}\!\!
\right) \left(\!\!
\begin{array}{cc}
  \dd^B\delta_{\alpha\beta} & 0 \\
  0 & \!\!\!\!\!\!-\dd^D \\
\end{array}\!\!
\right) \left(\!\!\begin{array}{cc}
  S_{0\,\beta\nu}^{\phantom{0\,}\Bb B} & \!\!\!S_{0\,\beta}^{\phantom{0\,}\Bb D} \\
  -S_{0\,\nu}^{\phantom{0\,}\Bb D} & \!\!\!-S^{\Db D}_0 \\
\end{array}\!\!
\right).
\]
A much more elegant description involves combining these into
five-dimensional fields but for simplicity we will not do so here.

\subsection{Effective propagator relations} 
\begin{figure}[h!]
\begin{center}
\includegraphics[height=1cm]{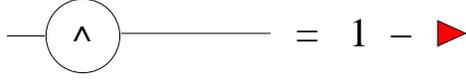}
\end{center}
\caption{Diagrammatic representation of the effective propagator
relations.}
\label{effprop}
\end{figure}

From the above flow
equations we determine the effective propagators and thus get a
relation of the same form as \eq{scaeffprop}:
\[
S_0^{CC} \Delta^C = 1,
\]
and relations analogous to \eq{Aeffprop}:
\[
S_{0\,\mu\ \nu}^{\phantom{0}A_1\!A_1} \Delta^1 = \delta_{\mu\nu} -
p_\mu p_\nu/p^2 \ins11{and} S_{0\,\mu\ \nu}^{\phantom{0}A_2\!A_2}
\Delta^2 = \delta_{\mu\nu} - p_\mu p_\nu/p^2.
\]
Once again, the effective propagators here are not really
propagators: multiplying the two-point vertex they give unity up
to terms that simplify via gauge transformations. For the $B$s and
$D$s one finds
\[
\left(\!\!\begin{array}{cc}
      S_{0\,\mu\alpha}^{\phantom{0\,}\Bb B} & \!\!\!S_{0\,\mu}^{\phantom{0\,}\Bb D} \\
      -S_{0\,\alpha}^{\phantom{0\,}\Bb D} & \!\!\!-S^{\Db D}_0 \\
      \end{array}\!\!\right) \left(\!\!\begin{array}{cc}
       \Delta^B\delta_{\alpha\nu} & 0 \\
       0 & \!\!\!\!\!\!-\Delta^D \\
      \end{array}\!\!\right) = \left(\!\!\begin{array}{cc}\delta_{\mu\nu} & 0\\ 0&1\\
      \end{array}\!\!\right) -
      \left(\!\!\begin{array}{c}f{p_\mu\over\Lambda^2}\\ g\\
      \end{array}\!\!\right)\left(p_\nu  \ 2 \right).
\]
This is unity up to a term which simplifies using spontaneously
broken supergauge transformations: the $(p_\nu\ 2)$ part operating
on a vertex can be simplified via broken Ward identities. Its
coefficient contains the cutoff dependent terms $f_p$ and $g_p$,
whose precise structure depends on the parameterisation we used
for the two-point vertices \cite{aprop}. We can unify all of these
relations diagrammatically as in \fig{effprop}. We write the
coefficient parts as a single unfilled arrow \ie as
``${\color{red}>}$'' and fill this arrow in when we include also
the terms that generate gauge transformations (null, $p_\nu$ and
$(p_\nu\ -2)$ for the $C$, $A$ and $B/D$ sectors respectively).
Then in all sectors, the two-point classical action (or seed
action) vertex attached to an effective propagator gives 1 minus a
term which simplifies on using Ward identities.

\subsection{`Na\"\i ve' Ward identities} 
\begin{figure}[h!]
\begin{center}
\includegraphics[height=2cm]{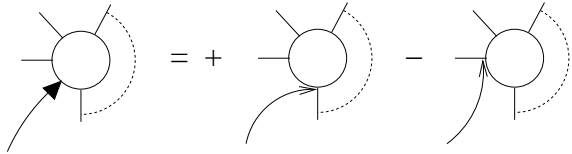}
\end{center}
\caption{Diagrammatic representation of the Ward identities. Only
one push-forward (counterclockwise arrow) and pull-back (clockwise
arrow) have been drawn. By symmetry they appear on all legs of the
vertex.}
\label{ward}
\end{figure}

To see what these Ward
identities look like, it is enough to consider the simple case in
pure $SU(N)$ gauge theory of a vertex of the form
\[
S = {\cdots +}\int S_{\mu_1\mu_2\cdots\mu_n}(p_1,p_2,\cdots,p_n)
    \,\tr\, A_{\mu_1}(p_1)A_{\mu_2}(p_2)\cdots A_{\mu_n}(p_n)
    \ \ {+\cdots}.
\]
This is because the diagrammatic representation turns out to be
the same for any vertex in all sectors \cite{thesis,6,7}.
Requiring that the action be invariant under infinitessimal gauge
transformations \eq{gaugetr}, \ie $\delta A_\mu =
\partial_\mu\omega-i[A_\mu,\omega]$, results in the so-called
na\"\i ve Ward identities:
\[
p_1^{\mu_1}S_{\mu_1\mu_2\cdots\mu_n}(p_1,p_2,\cdots,p_n)
     = \phantom{S_{\mu_1\mu_2\cdots\mu_n}(p_1,p_2,\cdots,p_n)}\]
\[   \phantom{S_{\mu_1\mu_2\cdots\mu_n}()}
     S_{\mu_2\cdots\mu_n}(p_1\!+\!p_2,p_3,\cdots,p_n) -
     S_{\mu_2\cdots\mu_n}(p_2,p_3,\cdots,p_n\!+\!p_1).
\]
They are called ``na\"\i ve'' because in the gauge fixed approach
one should really be using BRST invariance which leads to
complications and ultimately the Lee-Zinn-Justin identities
\cite{book,LZJ}. Here, because the gauge invariance is preserved
we get the above simple form. (The commutator part of the
non-Abelian gauge transformation provides the RHS which would just
be zero in an Abelian gauge theory.) In the diagrammatic
representation we see that the contraction of the term generating
gauge transformations results in the momentum of that leg being
`pushed forward' onto the following leg (with a plus sign) or
`pulled back' onto the previous leg (with a minus sign) as
illustrated in \fig{ward}.
In fact, since the vertex has been symmetrised over permutations
(as usual in Feynman diagrams), the pull-backs and push-forwards
operate on all legs. 

\section{Diagrammatic Method}
We now have all the elements necessary to describe the full
diagrammatic method. The flow equation has the same diagrammatic
form as before, \viz \fig{fAflow}. All we have to remember, in
fact most of the time we can forget, is that propagating between
vertices is now not just the gauge field $A^1$ but all the other
partners ($A^2$, $B$, $C$, $D$), which are necessary for
regularisation. This means only that internal legs (kernels or
effective propagators) must be interpreted as summed over these
flavours.

\begin{figure}[h!]
\begin{center}
\includegraphics[height=2.3cm]{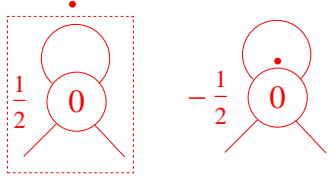}
\end{center}
\caption{Shifting the RG-time derivative. The differential is now
outside the integral in the first expression.}
\label{oneLoop-2}
\end{figure}

As a consequence the one-loop diagrams look the same as before,
\viz \fig{oneLoop}, however these diagrams converge, not only in
the infrared (as guaranteed by the Wilsonian formalism for smooth
cutoffs) but also now in the ultraviolet. Therefore, we can just
compute them. However we want to proceed without having to specify
the regularisation structure (cutoff functions, covariantization,
and seed action) precisely. This leaves us with very little room
to maneouver. The only worthwhile operation is to shift the
RG-time derivative in the first diagram by integration by parts,
resulting in \fig{oneLoop-2}.
Since the integral in the first diagram is a contribution to the
$\beta$ function, it is dimensionless. {\it A priori} the integral
can depend only upon $\Lambda$. However, since $\Lambda$ has
dimensions of mass and there is no other dimensionful parameter,
the integral is in fact independent of this and is thus a pure
number. Differentiating it gives zero.

\begin{figure}[h!]
\begin{center}
\includegraphics[height=3cm]{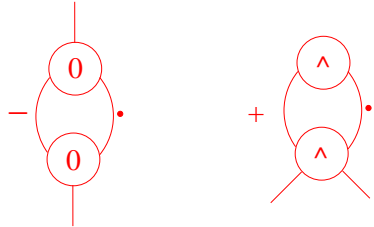}
\end{center}
\caption{Examples of diagrams obtained from expanding the second
diagram in \fig{oneLoop-2}, together with their combinatorics.}
\label{oneLoop-5}
\end{figure}

\begin{figure}[h!]
\begin{center}
\includegraphics[height=2cm]{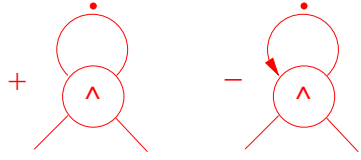}
\end{center}
\caption{The result of applying effective propagator relations to
the second diagram in \fig{oneLoop-5}.}
\label{oneLoop-7}
\end{figure}

\begin{figure}[h!]
\begin{center}
\includegraphics[height=2cm]{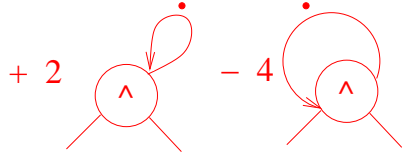}
\end{center}
\caption{The result of applying the Ward identities to the second
diagram in \fig{oneLoop-7}.}
\label{oneLoop-9}
\end{figure}

\begin{figure}[h!]
\begin{center}
\includegraphics[height=3cm]{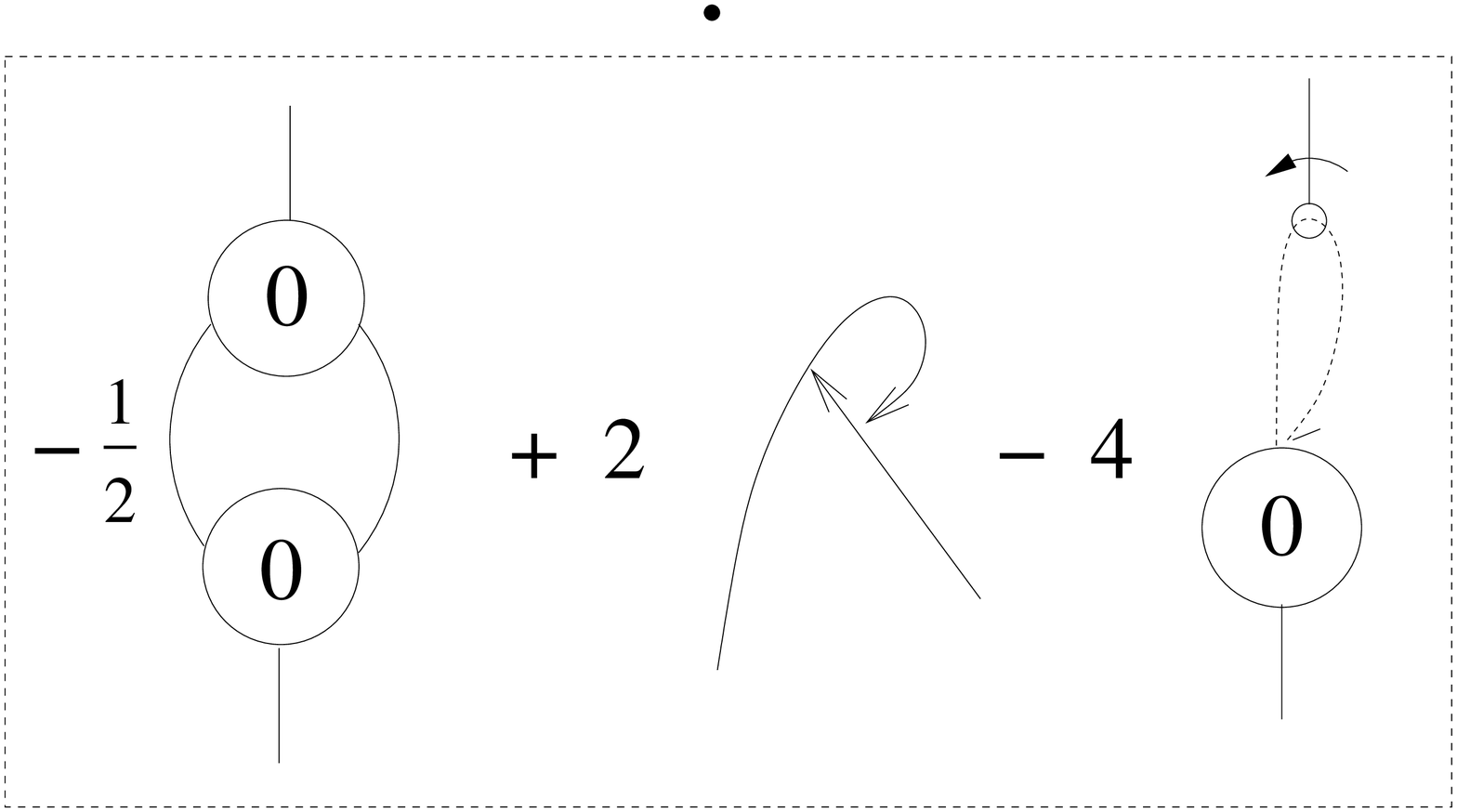}
\end{center}
\caption{The final diagrams contributing the one-loop $\beta$
function.}
\label{oneLoopT}
\end{figure}

We have thus effectively shifted the RG-time derivative onto the
classical vertex, as in the remaining diagram of \fig{oneLoop-2}.
We can now expand this flow using \fig{fA04flow}. This results in
a bunch of diagrams, corresponding to all possible ways of tieing
up two legs in the diagrams on the RHS of \fig{fA04flow}. Let us
concentrate on just two of these. The first diagram displayed in
\fig{oneLoop-5} corresponds to tieing up a top and a bottom leg
from the first diagram on the RHS of \fig{fA04flow}.
The second diagram in \fig{oneLoop-5}, the result of tieing up the
top and bottom legs of the third diagram on the RHS of
\fig{fA04flow}, can be processed further. We use the effective
propagator relations, \fig{effprop}. The result is
\fig{oneLoop-7}.
The first diagram in this figure is equal and opposite to the
second diagram in the original list, \fig{oneLoop}, so that now
two diagrams have disappeared from this list. The second diagram
in \fig{oneLoop-7} can be processed further by using the Ward
identities, \fig{ward}, pushing-forwards and pulling-backwards to
give \fig{oneLoop-9}.
We continue to proceed in the same way, for example the first
diagram of \fig{oneLoop-5} can be expressed as a total derivative
minus terms where the RG-time derivative hits the classical
three-point vertices. These can thus be expanded. Evaluating these
and others in the same way, we find that all diagrams cancel,
including the originals in \fig{oneLoop}, except for the set of
total derivative diagrams in \fig{oneLoopT} \cite{thesis,6}. 
(The third diagram is the result of expanding to $O(p^2)$, as
required from \eq{beta1}, and processing the resulting derivatives
with respect to $p^\lambda$ using differential Ward identities
\cite{thesis,6}.)

We saw in \fig{oneLoop-2} that the total RG-time derivative of a
diagram contributing to $\beta$ is zero because the integral can
only give a pure number. The only reason that these diagrams are
not also zero is that in the exchange of limits, placing the
differential with respect to RG-time outside the momentum
integral, we have introduced infrared divergences. In other words,
strictly speaking we were not allowed to exchange limits in this
way, however doing so has the advantage that we see very clearly
that the results for these integrals depend only on the massless
fields (\ie the $A^i$) propagating in the far infrared. For such
vanishing momenta, all the elements in the integral are then
determined by the renormalization conditions. That is why the
result is universal. It is easy to compute the resulting
integrals, and thus we obtain \cite{6,aprop}:
\[
{22N\over3(4\pi)^2}\,\Box_{\mu\nu}(p)+O(p^4).
\]
By comparing to our starting point, \eq{beta1}, we extract the
famous asymptotically free one-loop beta function for Yang-Mills.

The procedure generalises to two loops and beyond. The main new
feature is that subtraction diagrams need to be defined in order
to isolate the universal parts \cite{thesis,twoloops}. In this
way, we have confirmed that the universal two-loop beta function
is also reproduced by these methods.

\section{QCD}

Spinors transforming in the fundamental representation can be
incorporated in the framework, so we can compute in the physically
important theory of Quantum Chromodynamics, without fixing the
gauge \cite{qcd}. These fermions are the quarks; the gauged group
is $SU(3)$; $A^1$ is thus the gluon field.

The obvious thing to try, is to extend the quark spinor $\psi$ to
a fundamental representation $\Psi$ of $SU(3|3)$. However, gauge
invariance then fixes the interaction to be of the form $\Psi\A \!
\! \! /\Psi$ which violates the no-$\A^0$ invariance \eq{noA0}.
The only way out is to try a different embedding which does not
have this problem. In fact we can embed $\psi$ in the off-diagonal
part of a $3\times3$ supermatrix (transforming as a complexified
$U(3|3)$ adjoint field):
\[
\Psi = \left(%
\begin{array}{cc}
  \phi^1 & \psi \\
  \varrho & \phi^2 \\
\end{array}%
\right).
\]
In this way $\psi$ is fermionic by default. It transforms as the
fundamental under colour, $SU_1(3)$, and complex conjugate
fundamental under $SU_2(3)$. Thus we are forced into a situation
where $\Psi$ represents multiples of three flavours of quark. From
this point of view it is at least serendipitous that nature
decided that as well as three colours, there are also three
families. One super-spinor $\Psi_u$ can thus contain the up-type
quarks $u$, $c$ and $t$, and another super-spinor $\Psi_d$ hold
the down-type quarks $d$, $s$ and $b$. Of course at this stage the
family symmetry $SU_2(3)$ is gauged, moreover by the unphysical
field $A_2$. To give the quarks different masses, we need to break
it spontaneously (so that we keep the regularising properties of
$SU(3|3)$ at high energies).



(Although it is cute that families come in multiples of the number
of colours, there is no restriction to handling different numbers
of families since we can always send individual quark masses to
infinity at the end of the calculation.)

\section{Other applications}

We note that these ideas have been generalised, or rather
simplified, to the case of QED \cite{qed}, and the steps necessary
for computing general gauge invariant matrix elements have also
been fleshed out and incorporated in this general scheme, using
the example of the one-loop contribution to the Wilson loop
\cite{ops}. We were also able to use these ideas to make explicit
the cutoff in the AdS/CFT construction \cite{Nick}.

So far we have only discussed explicitly the perturbative
expansion of the flow equations, and concentrated on the
evaluation of the first two $\beta$ function coefficients, without
gauge fixing. The fact that this all works however, gives us
experience and confidence in applying the framework to the
non-perturbative regime. As we intimated in the introduction,
there is a wide variety of possible approximations that can be
applied.

Indeed within this framework, we have already studied the large
$N$ approximation to some extent in \cite{0,1}. The similarities
to string theory that this uncovered, together with the gauge (and
Poincar\'e) invariant cutoff definition in this framework, and
suggestions of exact RG from the AdS/CFT side \cite{verlinde},
provided the initial motivation to search for some connection with
the AdS/CFT construction.

We would also like to mention that it is possible to perform a
sort of Eguchi-Kawai reduction \cite{EK} of the equations, this
time in the continuum, casting the equations in an equivalent way
in terms of vertices that carry no momentum \cite{unpub}.

Of course one can also simply resort to truncations, a strategy
that has been very popular in the exact RG field
\cite{reviews,reviewst}.

We would like to finish by showing that under some very general
conditions, confinement is a natural consequence in this framework
\cite{unpub}. We then describe in some detail one further
possibility adapted to this observation, namely a strong
renormalized coupling expansion, which indeed could be combined
with any of the above approaches and could have more general
applicability to other exact RGs \cite{unpub}. As well as being a
desirable starting point for further approximation, it means that
one can test for confinement within this framework by instead
testing whether these general conditions hold true or not.

\section{Confinement}

From the perturbative $\beta$ function we know that $g(\Lambda)$
grows as $\Lambda$ gets smaller. Although we cannot trust
perturbation theory in this regime, the natural expectation is
that $g$ grows without limit. Indeed in any formulation where the
gauge invariance is manifest, including lattice gauge theory, the
gluon itself cannot have a mass gap. Intuitively, in this case
confinement only follows if the coupling diverges in the infrared.
(In the gauge fixed theory, arguments based on the Kugo-Ojima
criterion \cite{KO} actually come to the opposite conclusion: that
$g$ freezes out at some finite value. There is some evidence for
this in gauge fixed formulations \cite{ex}, however these
arguments depend by their very nature on the ghost fields and thus
have nothing to say about a manifestly gauge invariant approach.
In any case, {\it a priori} both pictures can be correct since
they refer to different non-perturbative definitions of $g$.)

Our first condition then is that $g(\Lambda)\to\infty$ as
$\Lambda$ decreases. In principle, we could find that $g$ diverges
at a finite critical value $\Lambda\to\Lambda_{cr}>0$ but in this
case we would have an obstruction to completing the computation in
this way, which requires that we integrate out all the modes, \ie
that we take the limit $\Lambda\to0$. It may be that a change of
variables will allow us to pass smoothly over $\Lambda_{cr}$, \ie
that the problem is a kind of `coordinate singularity'. If this
can be done in terms of some $SU(N|N)$ fields $\A$ and $\C$, it
amounts to a different choice for $\Psi$ in \eq{erg}, \ie a
different choice of exact RG.

Therefore we require that $g(\Lambda)\to\infty$, but only as
$\Lambda\to0$; we require this for at least one of the infinitely
many choices that we have for gauge invariant exact RGs, if we are
to find confinement in this framework.

Remember that $g$ is defined by the renormalization condition
\eq{g}, which tells us that the effective two-point $A^1$ vertex
has the form
\be
\label{gper}
S_{\mu\nu}(p) = \Box_{\mu\nu}(p)/g^2+O(p^4).
\ee
It follows that in this case, in the limit $\Lambda\to0$ we find
$S_{\mu\nu}(p)\sim O(p^4)$. This is a well-known signal of
confinement. Indeed the three dimensional Fourier transform of the
(gauge fixed) full effective propagator gives the effective
potential $V(r)$ between two colour charges. In the perturbative
regime one has for large $r$, the standard Coulomb potential, as
follows trivially from dimensional analysis:
\[
V(r) \sim \int {d^3p\over p^2}\, \e{i{\bf p}\!\cdot\!{\bf r}} \sim
{1\over r}.
\]
In this non-perturbative regime we have instead the famous
linearly rising form for large $r$, corresponding to a constant
Yang-Mills string tension and confinement:
\[
V(r) \sim \int {d^3p\over p^4}\, \e{i{\bf p}\!\cdot\!{\bf r}} \sim
r,
\]
as again follows trivially by dimensional analysis. In what
amounts to much the same computation, one can show that changing
the propagator from $1/p^2$ to $1/p^4$ changes the lowest order
contribution to the expectation value of a Wilson loop from
perimeter law to area law (another standard signal of
confinement).

We therefore come to the conclusion that in this manifestly gauge
invariant exact RG formalism, confinement occurs if and only if
$g(\Lambda)\to\infty$ as $\Lambda\to0$.

\section{Renormalized strong coupling expansion}
Although this observation is neat, we cannot use it as a basis for
computation in this regime because it only makes sense for
momentum $p\approx0$ (which is in fact all that is needed to
confirm the form of the potentials above for large $r$). The
reason is as follows. Since we are dealing with a continuum limit
the effective action must be in self-similar form (\cf the
introduction), which means that the only explicit scale in the
solution is $\Lambda$. This means that the expansion in \eq{gper}
is in $p^2/\Lambda^2$, and thus as $\Lambda\to0$ the approximation
$S_{\mu\nu}(p)\sim O(p^4)$ becomes valid only for vanishing
momenta.

We should expect that dimensional transmutation takes place so
that the scale in the problem is really set by
$\Lambda_{YM}$.\footnote{We restrict the discussion here to the
Yang-Mills we have been addressing, although it can be applied to
QCD also, where the corresponding scale is $\Lambda_{QCD}$.} In
this way, we ought to be able to address the case of non-vanishing
momenta.

We will show that we can do so within a new form of expansion in
the renormalized coupling, namely in $1/g^2$. We leave as an open
problem whether such an expansion actually exists. Indeed if such
an expansion of the effective action is substituted into the exact
RG \eq{gierg}, the equations (unsurprisingly) do not close, so its
existence can probably only be tested within some further
approximation.

(Of course strong coupling expansions have a long history
\cite{Wstrong} but these were in the bare coupling $g_0\sim
g(\Lambda_0)$. They are known to have a finite radius of
convergence, which is problematic, since by asymptotic freedom we
require $g_0\to0$ in the continuum limit.)

By gauge (and Poincar\'e) invariance and dimensions, the $A^1$
two-point vertex takes the form $S_{\mu\nu}(p) =
s(p,\Lambda)\,\Box_{\mu\nu}$, for some dimensionless function $s$.
By the requirement of self-similarity this function can be written
as $s(p/\Lambda,g)$, where $g(\Lambda)$ runs according to
$\Lambda\partial_\Lambda g = \beta(g)$.

We now assume that in the low energy regime, $S$ is analytic in
$1/g^2$ for large $g$, and therefore can be written as an
expansion in $1/g^2$. For $s$, this implies
\be
\label{sexp}
s = s_0(p/\Lambda) + {1\over g^2} s_1(p/\Lambda) + {1\over g^4}
s_2(p/\Lambda) + \cdots,
\ee
for some functions $s_i$. This will have the consequence that
$\beta$ itself has an analytic strong coupling expansion:
\[
\Lambda\partial_\Lambda1/g^2 = {\tilde\beta}(1/g^2) =
\tb0+\tb1/g^2+\tb2/g^4+\cdots.
\]

The flow $\Lambda\partial_\Lambda 1/g^2={\tilde\beta}$ can now be
solved in the regime of large $g$ with the result that if $g$ is
to diverge but only as $\Lambda\to0$, then $\tb0=0$ and $\tb1>0$,
and
\be
\label{gexp}
{1\over g^2} = \left({\Lambda\over\Lambda_{YM}}\right)^{\tb1}
+{\tb2\over\tb1}\left({\Lambda\over\Lambda_{YM}}\right)^{2\tb1}+\cdots,
\ee
where the unit coefficient for the first RHS term amounts to our
definition of $\Lambda_{YM}$ and the neglected terms are of order
$\left({\Lambda/\Lambda_{YM}}\right)^{3\tb1}$. (It is also
possible that $\tb1=0$ and $\tb2>0$, however in this case
$\Lambda_{YM}$ would indicate a maximum $\Lambda$ above which the
renormalized strong coupling expansion breaks down. We will not
investigate this interesting possibility further here.)

We assume that $S$ stands only for the part of the effective
action that obtains a finite limit as $\Lambda\to0$. (We will not
address how to isolate this part of the effective action from the
regularisation structure which of necessity diverges in this
limit. For the Polchinski effective action it would be everything
except the regularisation in the kinetic term. It is not so simple
here, however all expectation values of gauge invariant operators
and their correlators, are automatically isolated from these
divergent parts. They are introduced into the action by being
coupled via source terms \cite{1,ops}.)

Since $g$ is a function of $\Lambda/\Lambda_{YM}$, we can write
the limit $S|_{\Lambda=0}$ as an RG invariant action $S^{inv}$. To
illustrate with the two-point vertex $s(p/\Lambda,g)$, we write it
equivalently as the function ${\tilde
s}(p/\Lambda_{YM},\Lambda/\Lambda_{YM})$. Its limit, ${\tilde
s}(p/\Lambda_{YM},0)$, can instead be written as the $\Lambda$
independent function $s^{inv}(p/\Lambda,g)$.

If we now assume that $S^{inv}$ also has a strong coupling
expansion then, from
\[
0=\Lambda\partial_\Lambda S^{inv}= \Lambda\partial_\Lambda|_g
S^{inv}+{\tilde\beta}\,\partial_{1/g^2} S^{inv},
\]
we constrain the form of its expansion coefficients, and in
particular for $s^{inv}$ we find
\[
  s^{inv}_0 = \sigma_0,\qquad
  s^{inv}_1 = \sigma_1 \left({p\over\Lambda}\right)^{\tb1},\qquad
  s^{inv}_2 =
  \sigma_2\left({p\over\Lambda}\right)^{2\tb1}
  -\sigma_1{\tb2\over\tb1}\left({p\over\Lambda}\right)^{\tb1},
\]
and so on, where the $\sigma_i$ are numerical coefficients.
Combining with the expansion \eq{gexp}, we find the expected
dimensional transmutation:
\be
\label{sinvexp}
s^{inv} = \sigma_0
+\sigma_1\left({p\over\Lambda_{YM}}\right)^{\tb1}+\sigma_2\left({p\over\Lambda_{YM}}\right)^{2\tb1}
+\cdots.
\ee

We see that the assumption of a renormalized strong coupling
expansion and the existence of a limit as $\Lambda\to0$, implies
that this expansion turns into an expansion in small
$p/\Lambda_{YM}$. Since we have a mass gap, we should find that
the resulting expansion is analytic in $p$ (otherwise there would
be long-distance structure in position space). In this case we
would have to find that $\tb1$ is a positive even integer
(probably $\tb1=2$).

Subtracting $S^{inv}$ from $S$, leaves a remainder $R$ which thus
also has a strong coupling expansion:
\[
S = S^{inv}+R_0+R_1/g^2+R_2/g^4+\cdots.
\]
Since $S\to S^{inv}$ as $\Lambda\to0$, and since $g$ can be varied
independently of the $R_i$ (by varying $\Lambda_{YM}$), we must
have that $R_n/g^{2n}\to0$ as $\Lambda\to0$ at fixed finite $p$
and $\Lambda_{YM}$. From \eq{gexp}, this implies
$R_n\Lambda^{n\tb1}\to0$ as $\Lambda\to0$. Of course this bounds
above the amount by which $R_n$ can diverge as $\Lambda\to0$. For
the two-point vertex, let us call the remainder coefficients
$r_n(p/\Lambda)$. Then by dimensions, $r_n/p^{n\tb1}\to0$ as
$p\to\infty$. On the contrary, from \eq{sinvexp} and \eq{sexp}, we
have that the $s$ expansion coefficients have a unique finite
limit: $s_n/p^{n\tb1}\to \sigma_n/\Lambda^{n\tb1}$ as
$p\to\infty$. We therefore have an upper bound on the behaviour of
$r_n$ and $s_n$ as $p\to\infty$.

Recall that we are interested in the limit $\Lambda\to0$ at fixed
finite $p$. We have thus seen that the renormalized strong
coupling expansion is restricted to the regime $\Lambda<\!\!< p
<\!\!< \Lambda_{YM}$. We therefore cannot use the definition of
$g$ in \eq{gper}. If we can extract the coefficients (\eg the
$r_n$ and $s_n$ above) from the exact RG equations, then at finite
$\Lambda$ we can expect to find that they have a complicated form,
certainly not polynomial in $p/\Lambda$. From the above analysis
we see that they are bounded above by some power of $p/\Lambda$.
We therefore expect that these functions can be expanded as a
series in $\Lambda/p$. The natural generalisation of \eq{gper} to
this case is to define $1/g^2$ to be the momentum independent
coefficient\footnote{Remember that $s$ is the coefficient of
$\Box_{\mu\nu}$ in the effective two-point vertex.} in the
expansion of $s(p/\Lambda,g)$ in $\Lambda/p$. We emphasise however
that this is a different definition from the perturbative one in
\eq{gper}. There is no reason to expect them to
coincide.\footnote{This follows from simple analysis; compare for
example the $x$ independent terms in the small $x$ expansion, and
small $1/x$ expansion, of $1/(1+x)$.}

Since $\lim_{p/\Lambda\to\infty}s_0(p/\Lambda)=\sigma_0$, we have
from \eq{sexp} and this renormalization condition that
$\sigma_0=0$. Since $s\to s^{inv}$, we recover confinement from
\eq{sinvexp}, by the arguments of the previous section, providing
indeed that we find that $\tb1=2$.

\section*{Acknowledgement} TRM thanks the Fields Institute, and the Royal Society and
University of Helsinki for financial support to attend the
workshop and conference respectively.

\end{document}